\documentclass[sigconf,pbalance]{acmart}
\usepackage{lineno}
\usepackage{array}
\usepackage{booktabs,multirow}
\usepackage{makecell}
\usepackage{graphicx}
\usepackage{flushend}
\AtBeginDocument{%
  }

\makeatletter
\renewcommand{\footnotetextcopyrightpermission}[1]{%
  \footnotetext{%
    Permission to make digital or hard copies of all or part of this work for personal or classroom use is granted without fee provided that copies are not made or distributed for profit or commercial advantage and that copies bear this notice and the full citation on the first page. Copyrights for components of this work owned by others than the author(s) must be honored. Abstracting with credit is permitted. To copy otherwise, or republish, to post on servers or to redistribute to lists, requires prior specific permission and/or a fee. This paper has been accepted by ACM Multimedia 2026.%
  }%
}
\makeatother

\begin{document}

\title{Robust Coverless Linguistic Steganography via Sentence Embedding Space with Global Resynchronization}

\author{Lizhi Xiong}
\correspondingauthor
\orcid{0000-0003-1604-7690}
\affiliation{%
  \institution{Nanjing University of Information Science and Technology}
  \city{Nanjing}
  \country{China}
}\email{lzxiong16@163.com}

\author{Yuping Lu}
\orcid{0009-0008-0826-3397}
\affiliation{%
  \institution{Nanjing University of Information Science and Technology}
  \city{Nanjing}
  \country{China}}
\email{202412200704@nuist.edu.cn}

\author{Jun Li}
\orcid{0009-0007-3589-1068}
\affiliation{%
  \institution{Nanjing University of Information Science and Technology}
  \city{Nanjing}
  \country{China}}
\email{lijuun@yeah.net}

\author{Ziqiang Li}
\orcid{0000-0001-9484-2310}
\affiliation{%
  \institution{Nanjing University of Information Science and Technology}
  \city{Nanjing}
  \country{China}}
\email{iceli@mail.ustc.edu.cn}

\author{Zhangjie Fu}
\orcid{0000-0002-4363-2521}
\affiliation{%
  \institution{Nanjing University of Information Science and Technology}
  \city{Nanjing}
  \country{China}}
\email{wwwfzj@126.com}

\renewcommand{\shortauthors}{Lizhi Xiong, Yuping Lu, Jun Li, Ziqiang Li, \& Zhangjie Fu}

\begin{abstract}
Linguistic steganography enables covert communication through natural language. Existing methods heavily rely on token-level operations and struggle to maintain reliability under word- and sentence-level textual perturbations. Moreover, variable-length coding-based schemes are highly susceptible to bit-slippage under minor disturbances, as perturbations cause desynchronization between embedded and extracted bit sequences.
To address these issues, we propose a robust coverless steganographic framework that operates in the sentence embedding space rather than the token space. Specifically, secret messages are encoded as hierarchical clustering paths in the sentence embedding space, which enhances decoding stability against word- and sentence-level textual perturbations. To tackle the bit-slippage problem, we introduce a Global Resynchronization Mechanism (GRM) that reframes variable-length bitstreams as discrete symbols anchored to semantic subspaces, decoupling local embedding failures from global message recovery.
Experimental results demonstrate that under word- and sentence-level perturbations, our approach achieves substantial improvements in robustness, while maintaining effective embedding capacity and exhibiting strong resistance to statistical analysis.
\end{abstract}




\maketitle
\pagestyle{plain}

\section{Introduction}
With the advancement and development of social networks, today's social media platforms increasingly rely on automated systems for large-scale and strict information regulation to monitor, filter, and shape public opinion. Under the circumstances, users are turning to steganography as a covert means of conveying sensitive information, aiming to protect personal privacy. Steganography is a technique that embeds secret information into seemingly innocuous carriers to enable covert communication without arousing suspicion. In the age of digital steganography, people use texts, images~\cite{wu2016separable}, videos~\cite{chen2019adaptive}, and audios~\cite{wu2020audio} as media to transmit secret messages. As the most fundamental medium of human communication, text is generated, transmitted, and consumed at massive scale across social media platforms every day, providing a natural and abundant cover for information hiding. Consequently, embedding secret information in text has become an important and active branch of steganography research.

\begin{figure}[tb]
\centering
\includegraphics[width=0.4\textwidth,height=4.5cm]{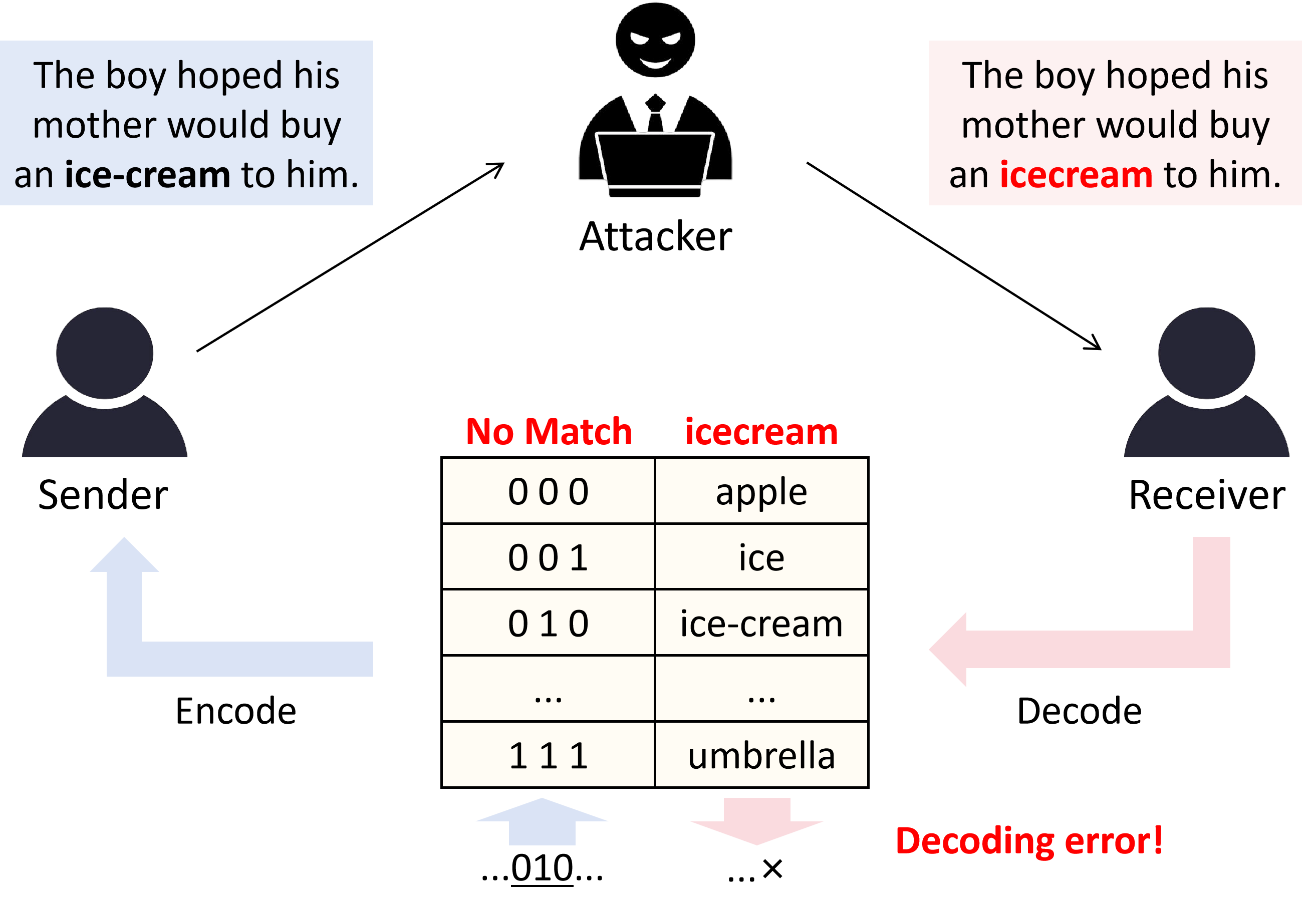}
\caption{An example of decoding error due to attacked token not in the vocabulary.}
\label{attack}
\end{figure}

Existing steganography methods are classified into modification-based, retrieval-based and generative steganography. In the first two methods, secret messages are conveyed by altering or selecting words~\cite{yi2022alisa} or sentences~\cite{wilson2016avoiding} according to predefined coding rules. Such methods are inherently fragile due to the sensitivity of the characters. Even slight deletions or replacements may break the shared dictionary/codebook correspondence, leading to incorrect decoding or complete extraction failure once the edited content falls outside the shared dictionary. 
Subsequently, generative linguistic steganography~\cite{yang2018rnn,zhou2021linguistic} based on Autoregressive Models (ARMs)~\cite{touvron2023llama} gained popularity due to fluent and natural generation. In these methods, secret information is embedded by sampling from context-conditioned token probability distributions; however, their strong sequential dependency makes them extremely fragile, as tampering with a single token can propagate errors to subsequent tokens and cause irreversible information loss, especially when attacks introduce error contents or segmentation ambiguity~\cite{qi2024provably}. As shown in Figure~\ref{attack}, for methods that use characters or tokens as steganographic channels, even minor perturbations can cause decoding to fail. Therefore, existing text steganography methods generally suffer from insufficient robustness when facing realistic text attacks, fundamentally limiting their practical applicability.

To address the problem, later robust steganographic methods~\cite{guo2025secc,pang2025winstega,qi2025stead} introduced some defense mechanisms at the token-level to enable partial recovery. Nevertheless, these defense mechanisms focus only on token-level perturbations while ignoring word- and sentence-level attacks. In practice, real-world adversaries are unaware of the steganographic system's segmentation rules and tend to operate at the word- or sentence-level through deletion, insertion anywhere or paraphrasing. Consequently, while token-level defenses can improve robustness against token-boundary-aligned perturbations, robustness under practical word- and sentence-level rewriting remains a fundamental and unresolved challenge for current text steganography frameworks.

To address these challenges, we propose a robust coverless linguistic steganography framework in the sentence embedding space. Specifically, a shared sentence encoder maps sentences into a unified semantic space, which is partitioned into multiple codable semantic subspaces for information hiding. We cluster the semantic space in a hierarchical manner to reduce the computational overhead associated with the clustering required for high-capacity. Secret information is conveyed through subspace selection and sentence sampling within the selected subspace. To ensure the security of subspace selection, a provably secure coding-and-sampling strategy is adopted to selectively encode only a subset of semantic subspaces, so that the resulting sampling distribution remains consistent with the underlying natural distribution. However, this strategy is inherently variable-length, and thus a subspace prediction error at the receiver may lead to a local mismatch between the extracted and embedded message lengths, causing bit-slippage across subsequent bits. To address this issue, we organize the hidden information as discrete symbols aligned with the partitioning structure of the semantic space and integrate LT codes for global recovery, thereby localizing decoding errors and improving robustness against text perturbations.

Our contributions can be summarized as follows:
\begin{itemize}
    \item We propose a novel coverless linguistic steganography framework that shifts the embedding operation from fragile token sequences to a robust, high-dimensional sentence embedding space. By leveraging the approximate semantic invariance of sentence embeddings, our method inherently absorbs word- and sentence-level textual perturbations.
    \item We propose a \textbf{Hierarchical Clustering Mechanism (HCM)} that cluster the semantic space into codable subspaces efficiently. By recursive pruning and hierarchical mapping, HCM achieves an optimal balance between embedding capacity and computational efficiency, reducing the complexity from exponential to linear.
    \item To resolve the critical issue of bit-slippage in variable-length provably secure encoding, we introduce a \textbf{Global Resynchronization Mechanism (GRM)}. By reframing local desynchronization as symbol erasures and integrating LT codes for global recovery, GRM decouples local embedding failures from global message integrity.
    \item Extensive experiments demonstrate that our method achieves strong robustness and effective embedding capacity under word- and sentence-level textual attacks.
\end{itemize}

\section{Related Work}

\subsection{Retrieval Linguistic Steganography}

Retrieval steganography requires the sender and receiver to share a large dataset and a set of coding rules, whose advantage is that the transmitted carrier remains natural without any modification. The sender selects a group of elements that can be mapped to the secret messages and sends them to the receiver to realize the transmission of secret information. For example, ~\cite{zhang2017coverless} established the mapping relationship through the word-frequency characteristics in the corpus; ~\cite{long2018text} encoded text according to the distance in semantic space; ~\cite{wang2019coverless} built a search tree according to the character structure; ~\cite{hu2020mm} jointly realized information hiding through a database combining graphics and text. Retrieval linguistic steganography is a type of coverless steganography. Current retrieval-based text steganography methods have not systematically investigated steganographic robustness, which needs to be further explored. These methods rely on ARMs and hardly change the probability distribution of the generated words or semantics to ensure security when embedding secret information.

\subsection{Provably Secure Linguistic Steganography}

In recent years, researchers have proposed a variety of provably secure linguistic steganography methods based on generative language models. These methods are dedicated to designing message-embedding algorithms that are indistinguishable from the conventional text-generation process. ADG~\cite{zhang2021provably} adaptively and dynamically grouped tokens based on the probability generated by the pre-trained language model and recursively embeds secret information. Meteor~\cite{kaptchuk2021meteor} used the random prefix shared by the generation model to embed the message into the sampling random number and masked it at once, dynamically adapted to entropy changes, and takes both security and efficiency into account. Discop~\cite{ding2023discop} proposed the ``distributed copies'' technique, which uses the index value of these ``copies'' to express secret information during the generation process and realizes practical steganography with provable security and high capacity. ~\cite{pan2025rethinking} made two major improvements to the prefix-encoding method—quantitative fine-tuning and distributed coupling—to repair security and capacity defects.

\subsection{Robust Linguistic Steganography}

In recent years, researchers began to pay attention to the problem of message loss caused by segmentation ambiguity, and designed corresponding countermeasures. SECC~\cite{guo2025secc} overlaid an ECC layer similar to LT codes~\cite{luby2002lt} on the conventional generative steganography pipeline to improve character-level erasure resistance; however, it didn't mitigate the original extraction-error problem. To avoid altering the probability distribution of candidate words, Syncpool~\cite{qi2024provably} merged tokens that share a prefix into an ``ambiguity pool'' and used a shared encrypted pseudo-random generator to ensure consistent selection, yet the secret information remained sensitive to vocabulary shifts.

Active textual attacks—such as deletion, insertion, and tampering—have also been studied. GTSD~\cite{wu2024gtsd} employed a diffusion model to generate candidate texts and embeds information via prompt-and-batch mapping, achieving local robustness. Winstega~\cite{pang2025winstega} sacrificed capacity for anti-editing capability by using a sliding window and entropy-threshold-based discontinuous embedding during inference. STEAD~\cite{qi2025stead} embedded duplicate codes in parallel at multiple locations within a single denoising step of a diffusion model and coordinated with neighborhood-search alignment to realize provably secure steganography resistant to insertion and deletion. These approaches achieved stable extraction under their respective token-level active attacks.

\subsection{Sentence Embedding Encoder}

Sentence embeddings are obtained by encoding text sentences with a sentence embedding encoder. After sufficient training, embeddings of semantically similar sentences exhibit higher similarity and smaller distance. Semantic similarity is computed by:
\begin{equation}
\mathrm{Sim}(A, B)
= \frac{A \cdot B}{\|A\| \, \|B\|}
= \frac{\sum_{i=1}^{n} A_i B_i}
{\sqrt{\sum_{i=1}^{n} A_i^2} \, \sqrt{\sum_{i=1}^{n} B_i^2}},
\label{eq:cosine_similarity}
\end{equation}
where $A$ and $B$ denote the vector representations of the sentences, and $A_i$ and $B_i$ are their $i$-th components.

Owing to different training parameters and embedding dimensions, each encoder exhibits distinct characteristics. For instance, the static encoder~\cite{mikolov2013efficient}, trained with modest parameters, yields a relatively flat and uniform semantic space suitable for small models, but its advantage is speed. In contrast, encoders based on large models~\cite{chen2024bge} produce a more comprehensive and high-dimensional semantic space at the cost of increased time and memory, yet offer stronger text-processing capabilities. Due to the strong similarity-matching ability, sentence embeddings are widely employed in downstream tasks such as retrieval~\cite{karpukhin2020dense}, clustering~\cite{su2021whitening}, etc.

\section{Proposed Method}

\subsection{Deployment Assumption and Threat Model}

The proposed framework focuses on a symmetric key steganography system: the sender and receiver need to share the same dataset, sentence embedding encoder, LT Encoding Symbol Identifiers (ESI) and pseudo-random number key to ensure the construction of the same sentence-level semantic space and the success of extraction. The adversary can apply semantic-preserving perturbations but has no access to shared secrets, aiming to disrupt communication without altering meaning. Consider a Probabilistic Polynomial-Time (PPT) adversary operating in a highly regulated environment whose duty is to detect and eliminate covert communication that may be hidden in public texts. The adversary intercepts any text transmitted over public channels and possesses two capabilities:

\noindent\textbf{Detection capability.} The adversary can apply steganalysis techniques to determine whether the suspicious texts may contain secret information. This corresponds to the security requirement.

\noindent\textbf{Tampering capability.} The adversary can tamper a text $s$ into $s'$, but within limits: the semantic content of $s$ remain unchanged. Let textual attack be a transformation operator: $\mathcal{A} : s \rightarrow {s}'$. This transformation satisfies approximate semantic invariance:
\begin{equation}
\left\|\mathbf{v}({s}') - \mathbf{v}(s)\right\|_2\le \epsilon,
\label{tamper}
\end{equation}
where $\mathbf{v}(s)$ denotes the semantic vector of $s$ and $\epsilon$ is the tampering threshold. This corresponds to the robustness requirement of steganography.

\subsection{Overview}

\begin{figure*}[t]
\centering
\includegraphics[width=1.0\textwidth, height=6.18cm]{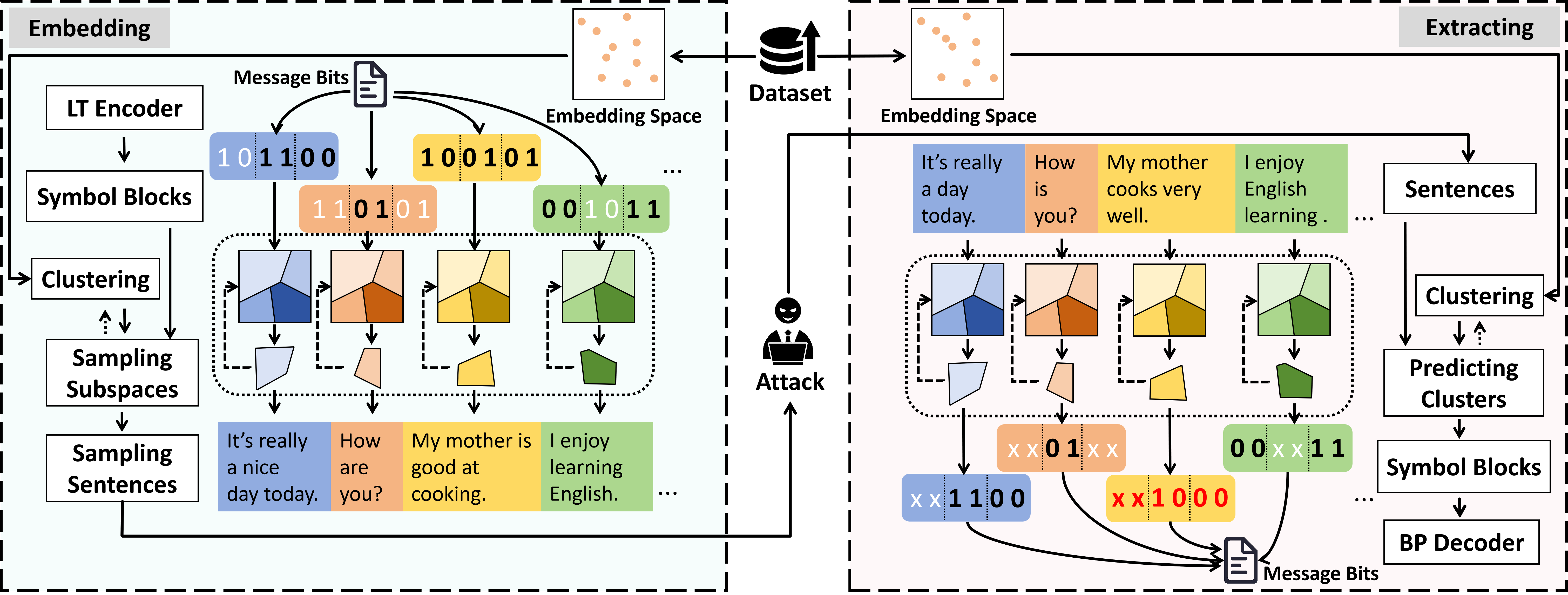}
\caption{An overview of the proposed stegosystem. Embedding and extraction are mutually inverse processes.}
\label{framework}
\end{figure*}

In this section, we introduce the proposed steganography framework. First of all, we introduce how to construct and partition semantic space and how to improve the efficiency. Then, we will describe the security application of SparSamp sampling method in this framework, and point out its limitations in robustness. On this basis, we employ a Global Resynchronization Mechanism (GRM) to improve it.

Figure~\ref{framework} shows the proposed stegosystem. In the embedding phase, we cluster the sentence embedding spaces mapped by the dataset, and code and sample the divided semantic subspaces at multiple levels. At the same time, we encode the secret messages to be sent into symbol blocks, and embed them into each subspace layer in units of symbol blocks. After the iteration, we randomly select sentence samples in the selected subspaces and send them as steganography carriers. Text attacks may occur during transmission. In the extraction phase, we use the same sentence encoder as in the embedding phase to reconstruct the semantic space of the shared dataset, predict the semantic subspace of each sentence embedding in the steganography texts, reproduce the same hierarchical clustering results as the embedding phase, and decode the symbol blocks according to the sampling process. Finally, we use Belief Propagation (BP) decoder to accurately recover the original secret messages according to the collected symbol blocks.

\subsection{Construction of Semantic Space}

\textbf{Robust Subspace Construction.} The core of our robust framework lies in the structural stability inherent in the high-dimensional sentence manifold. High-quality sentence embeddings, generated by pre-trained encoders, exhibit a crucial property of semantic invariance: sentences with minor textual perturbations tend to cluster within a constrained neighborhood in the semantic space $V_S$. By applying clustering algorithms to partition $V_S$ into a set of disjoint subspaces—rather than arbitrary partitioning—we capitalize on the natural distribution of the data to minimize intra-cluster variance. Due to the sparsity of high-dimensional spaces, most samples are distributed near the centroid. Consequently, we can establish a stable mapping between secret information and regional semantic characteristics. In this schema, we select the semantic subspaces as the fundamental steganographic units rather than individual tokens or specific sentence instances. Since the secret bits are tied to the subspace identity rather than the precise coordinates of a single vector, the system can tolerate significant word-level or sentence-level attacks. As long as the perturbed stego sentence ${s}'$ satisfies the semantic proximity constraint Eq.~\ref{tamper} and remains within the original cluster boundary, the receiver can accurately identify the corresponding subspace and retrieve the secret message. This structure-level design effectively decouples the hidden information from fragile textual surface forms, thereby providing intrinsic robustness against various practical attacks.

We use k-means algorithm to cluster a semantic space $V_{S}$. Given the number of clusters $k$ and pseudo-random seed $r_c$, k-means clustering can be formalized as the following optimization problem:
\begin{equation}
\begin{split}
\{\boldsymbol{\mu}_1,\dots,\boldsymbol{\mu}_k\}
&=
\arg\min_{\{\boldsymbol{\mu}_j\}_{j=1}^k}
\sum_{i=1}^{N}
\min_{1 \le j \le k}
\left\|
\mathbf{v}_i - \boldsymbol{\mu}_j
\right\|_2^2, \\
&\text{s.t.}\quad
\boldsymbol{\mu}_j^{(0)} \sim \mathcal{P}(r_c),
\end{split}
\label{kmeans}
\end{equation}
where ${\mu}_j$ is the $j$-th cluster center, ${\mu}_j^{(0)}$ is the initial centroid initialized by the pseudo-random number $r_c$, and $\mathcal{P}(r_c)$ is the initialization distribution determined by the random seed $r_c$. 

After clustering, the semantic space $V_S$ is divided into $k$ disjoint semantic subspaces. To ensure that each semantic subspace has sufficient stability, we test different values of $k$ to determine the appropriate threshold. When the optimization convergence time is too long or the subspaces are too small to be robust after clustering, we consider $k$ to be too large. When the value of $k$ is set too small, although there is no concern about the stability of the subspace, the embedding capacity becomes too limited. Since both sides of the communication share the same $r_c$, the clustering function is determined:
\begin{equation}
\mathcal{C}_{r_c} : V_{S} \rightarrow \{V_1, V_2, \dots, V_k\},
\label{map}
\end{equation}
thereby ensuring that the sender and receiver obtain consistent semantic subspace partition results.

\noindent{\textbf{Hierarchical Clustering Mechanism.}} If a high steganography capacity is required, a large dataset must be used to construct a richer semantic space, thereby enabling the partitioning of more semantic subspaces. As shown in Eq.~\ref{kmeans}, the time complexity of the k-means algorithm for $N$ samples is $O(N\cdot k)$. Assuming that a single sentence is to embed $b$ secret bits, Shannon's source-coding principle~\cite{shannon1948mathematical} dictates that $b$ secret bits requires at least $2^{b}$ distinguishable states. Consequently, the number of sub-clusters k must scale as:
\begin{equation}
k\propto 2^{b}.
\label{scale}
\end{equation}
Then the time complexity of the algorithm is $O(N\cdot 2^b)$. Therefore, when clustering large semantic spaces, increases in the number of samples $N$ and in the number of secret bits $b$ incur significant computational overhead.

To address this problem, we design a Hierarchical Clustering Mechanism (HCM) that embeds $b$ secret bits across $l$ layers, with each layer carrying $\bar{b}$ bits. In each layer $i$, the encoding process selects a specific cluster $V^{(i)}_{hit}$ as the ``hit'' subspace based on the secret bits and designates it as the semantic space $V^{(i+1)}$ for the next layer. This cycle repeats $l$ times.

The time complexity of this algorithm is $l\cdot O(N\cdot2^{\bar{b}})$. Based on the actual clustering efficiency, we set $\bar{b}$ to 2 or 3. Then the complexity of each layer of clustering can be controlled and stable, and the overall time complexity can be regarded as linear complexity $O(N\cdot l)$. By concentrating computational overhead solely on the steganographic path, this method significantly improves the overall efficiency of the framework and effectively performs recursive pruning of irrelevant branches in the semantic space.

\subsection{Global Resynchronization}

\noindent{\textbf{Subspace Sampling.}} During clustering, the discrete distribution of samples makes it impossible to partition the semantic space perfectly evenly, so the number of samples in each subspace is usually different. If the subspaces $\{V_1, V_2,..., V_k\}$ are encoded in simple sequential order, the secret-driven selection will bias the semantic distribution of the output sentences away from the natural distribution, introducing security risks. To eliminate this bias, we adopt the SparSamp~\cite{wang2025sparsamp} approach—message-driven pseudo-random number—for subspace sampling.

The construction of the semantic space and the procedure of HCM remain unchanged. Let the subspaces obtained after the $t$-th layer partition driven by the pseudo-random number $r_c$ be $\mathcal{C}_{r_c}(V^{(t)})=\{V_1^{(t)}, V_2^{(t)},..., V_k^{(t)}\}$, where $V^{(t)}=\sum_{i=1}^{k} V_i^{(t)}$. The cumulative probability distribution of the samples is then:
\begin{equation}
F^{(t)}(j)=\sum_{i=1}^{j} \frac{|V_i^{(t)}|}{|V^{(t)}|},
\end{equation}
where $j\in\{0,1,\dots,k\}$. If $b$ secret bits $m(t) \in \{0,1\}^b$ are to be embedded in this layer, we partition the probability interval $[0,1)$ uniformly with the minimum granularity $\delta=1/2^b$, yielding $2^b$ equally wide intervals whose starting positions are $x_i=i \cdot \delta$, where $i \in \{0,1,..., 2^b-1\}$. To simulate real random sampling and ensure reproducibility, we introduce a pseudo-random number $r_\Delta \in [0,1)$ derived from the shared key and construct the pseudo-random position:
\begin{equation}
r(i)=(x_i+r_\Delta)\  mod \ 1,
\end{equation}
where $i\in\{0,1,\dots,2^b-1\}$. Let $i^*=dec(m(t))$  be the decimal value of the secret bits. The subspace selected for the next layer is given by the inverse CDF:
\begin{equation}
V^{(t+1)}=V_{j^*}^{(t)},
\end{equation}
\begin{equation}
j^*=\arg\!\min_{j}\bigl\{F^{(t)}(j)\ge r(i^*)\bigr\}.
\end{equation}

If there are two or more pseudo-random positions corresponding to the target subspace $V^{(t+1)}$, no message is embedded in this layer. This phenomenon is known as a message conflict.

\begin{figure}[t]
\centering
\includegraphics[width=0.4\textwidth,height=6.18cm]{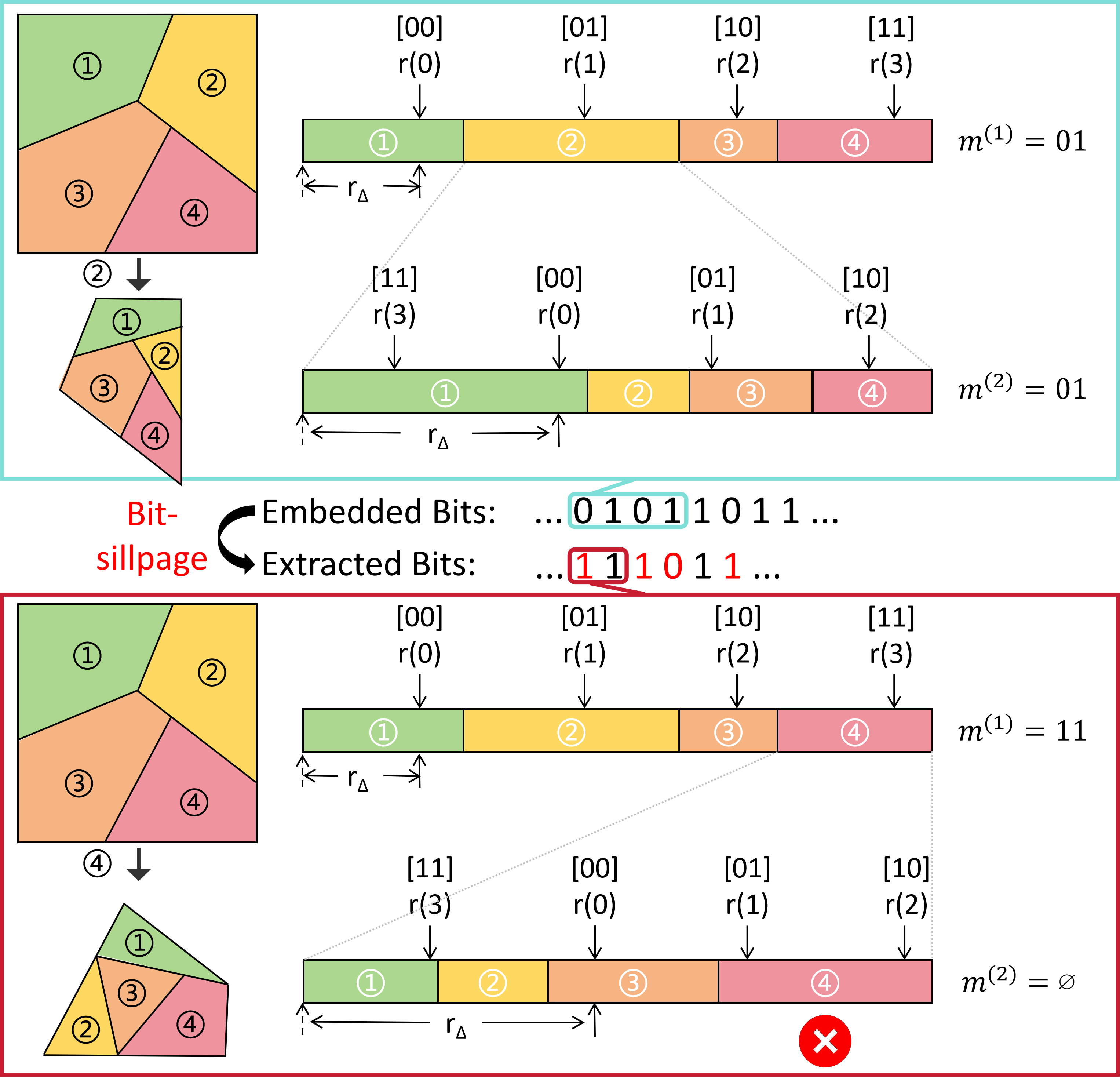}
\caption{An example of bit-slippage. Due to a local decoding error, the subsequent extracted bit sequence is out of sync with the embedded bits.}
\label{bit-slippage}
\end{figure}

\noindent{\textbf{Global Resynchronization Mechanism.}} After being subjected to a perturbation, there is a small probability that the receiver will incorrectly predict the cluster to which a sentence belongs, thereby extracting an incorrect secret message. More seriously, however, because the internal spatial structures of different clusters vary, the frequency of message conflicts also differs. This may cause the length of the bits extracted from that sentence to differ from the original bits, leading to misalignment in the subsequent bit sequence and significantly increasing the error rate. Figure~\ref{bit-slippage} illustrates an example of bit-slippage.

\begin{figure}[t]
\centering
\includegraphics[width=0.4\textwidth,height=4.8cm]{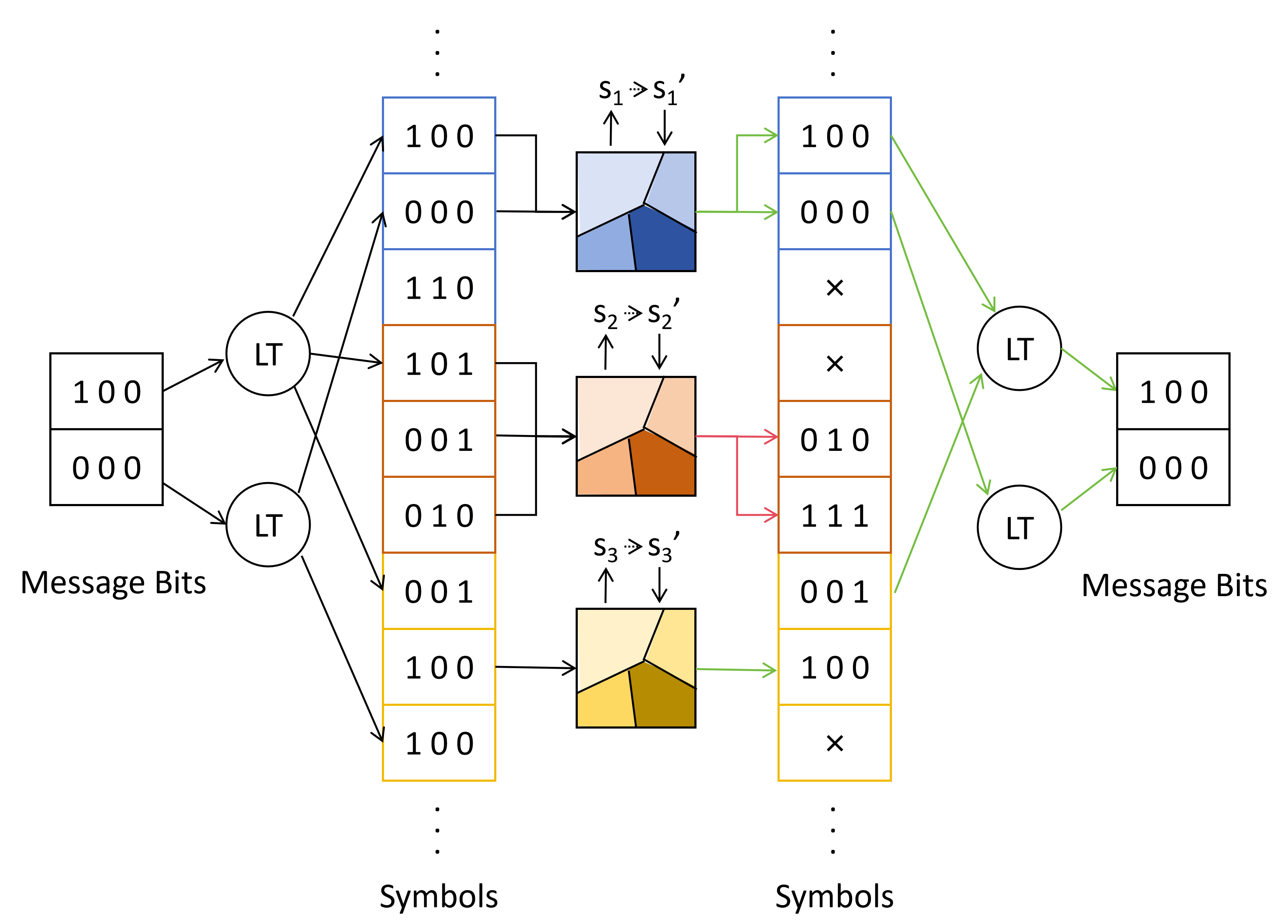}
\caption{An example of GRM. The green arrow indicates that the clustering is correct, and the red arrow indicates that the clustering is wrong.}
\label{ecc}
\end{figure}

To address this issue, we propose a Global Resynchronization Mechanism (GRM) based on LT codes to dynamically align the extracted bit positions and decouple local extraction failures from global message recovery. 

During embedding, we let the length of each symbol block of the LT codes be fixed to the single-layer embedding capacity $b$ and standardize the number of layers for each sentence clustering to $l$. The LT encoder can generate an unbounded checksum stream $C=\{c_1,c_2,\ldots\}$, where each symbol $c_i \in \{0,1\}^b$ represents a complete LT-encoded symbol block. Each layer carries exactly at most one symbol block. If the subspace sampling at layer $j$ of sentence $i$ does not trigger a conflict, the corresponding symbol block $B_{ib+j}$ is embedded into that layer. Otherwise, the layer is skipped and the symbol block $B_{(i-1)l+j}$ is not embedded. For the $i$-th sentence, the symbol blocks are preassigned to the interval $[(i-1)l+1,\, il]$ globally, where missing symbol blocks are treated as natural erasures.

During extracting, the same decoding procedure is applied to the received sentence $s'$: If the subspace sampling at layer $j$ of sentence $i$ does not trigger a conflict, the corresponding symbol block $B'_{ib+j}$ is extracted; Otherwise, the layer is skipped and $B'_{ib+j}$ is none. Due to clustering errors or message conflicts, symbol blocks may not be successfully recovered from sentence $s'$ that matches the one from the embedding stage. All extracted symbol blocks are subsequently reorganized into a global verification stream $C'$ according to their predefined ESIs. Once the total number of collected symbols exceeds the LT decoding threshold, the original secret message can be reconstructed in a single decoding step. In this manner, local layer-level deviations are isolated at the sentence level, preventing error propagation across bits and significantly enhancing the robustness of the steganographic system.

Figure~\ref{ecc} illustrates an example of GRM when $b=3$ and $l=3$. After being attacked, the sentence $s_2'$ is unfortunately divided into the wrong subspace, and the involved symbol blocks is wrongly extracted. Because LT codes with a certain degree of redundancy possess inherent robustness, the correct message can still be recovered even if some symbol blocks are missing.

\section{Experiment}

\subsection{Implementation Details}
\noindent{\textbf{Setup.}} We select the general text-embedding model Sentence-T5-Large~\cite{ni2022sentence} optimized by T5 model as the sentence-vector encoder. We retrieve and generate texts on PersonaChat~\cite{zhang2018personalizing} corpus. For HCM, we set the number of layers $l$ to 3 and the number of clusters per layer $k$ to 8. For decoding, we use the BP decoding algorithm.

\noindent{\textbf{Baselines.}} We select AC~\cite{ziegler2019neural}, ADG~\cite{zhang2021provably}, Discop~\cite{ding2023discop}, SECC~\cite{guo2025secc}, STEAD~\cite{qi2025stead} and FSBTS~\cite{yang2025novel} as the baseline methods. STEAD adopts Dream-7B~\cite{ye2025dream} and the rest adopt GPT-2~\cite{radford2019language}. FSBTS uses openConcepts~\cite{zhang2021alicg} as its dataset, while other methods randomly select texts from the PersonaChat corpus as inputs for text generation tasks.

\subsection{Metrics}

\noindent{\textbf{Robustness.}} In order to demonstrate the robustness, we test the extraction error rate $P_E$ of stegos after five types of textual attacks. In the actual extraction process, the damage of secret information is mainly manifested in two physical forms: fragment loss and extraction error. According to information theory, the mutual information provided by the completely lost bits is 0, which is equivalent to the random guess of equal probability (i.e. the error rate is 0.5) at the receiver in the binary channel. Therefore, we use the full probability formula to convert the bit loss rate (BLR) of the extracted message and the bit error rate (BER) of the extracted message into the equivalent comprehensive bit error rate $P_E$:
\begin{equation}
P_E=(1-BLR)*BER+0.5*BLR.
\end{equation}

The types of text attacks we use include: deletion, replacement, insertion, swap and paraphrase. We use Parrot\_Paraphraser~\cite{prithivida2021parrot} model to implement the paraphrase operation. Among them, paraphrase belongs to sentence-level attacks, and other attack types belong to word-level attacks. For word-level attacks, we set the ratio of the attacked words to the total number of words in a sample as $\alpha$, which can be regarded as the attack intensity.

\noindent{\textbf{Quality of Stegos.}} We use Perplexity (PPL) of the model DialoGPT-medium~\cite{zhang2020dialogpt} as the index to evaluate the quality of steganography. PPL measures the instability of an LLM in predicting text sequences and represents the fluency of texts. The smaller the PPL, the smoother the stegos and the better the concealment. Since the dataset of FSBTS is fixed, it is not included in the evaluation.

\noindent{\textbf{Embedding Capacity.}} The ultimate transmission rate for reliable, error-free communication over a memoryless binary channel is bounded by the channel capacity~\cite{shannon1948mathematical}. The channel uncertainty induced by the composite error rate $P_E$ is formulated by the binary entropy function $H_2(P_E)=-P_E\log_2(P_E)-(1-P_E)\log_2(1-P_E)$. At the theoretical limit, $H_2(P_E)$ represents the minimum proportion of redundant parity bits inherently required to correct channel noise. Conversely, the term $[1-H_2(P_E)]$ represents the capacity factor, which is the fraction genuinely available for carrying the useful payload after deducting the error-correction overhead. Consequently, the actual Effective Embedding Rate (EER) is computed as
$\mathrm{EER} = (1-H_2(P_E))*\mathrm{ER}$, where ER denotes the raw embedding capacity. As $P_E$ approaches 50\% and $[1-H_2(P_E)]$ approaches 0, more redundant codes need to be introduced, resulting in less EER. We conducted tests with $\alpha=5$, as a representative value for moderate intensity. Under this condition, all methods functioned properly, and there were significant differences in terms of robustness. Through numerous experiments, we have found that the relevant conclusions hold true for other values of $\alpha$ as well.

\begin{table}[t]
    \centering
    \caption{Robustness against paraphrase attack.}
    \label{para-robust}
    \resizebox{1\linewidth}{!}{
    \begin{tabular}{lccccccc}
        \toprule[1.5pt]
            Method & AC & ADG & SECC & Discop & STEAD & FSBTS & Ours \\
            \midrule
            BER & 0.6145 & 0.3933 & 0.5500 & - & - &0.1090 & 0.0038 \\ 
            BLR & 0	& 0	& 0	& 1	& 1	& 0.2681 & 0
            \\
            $P_E$ & 0.6145	& 0.3933 & 0.5500 & 0.5000 & 0.5000 & 0.2138 & \textbf{0.0038} \\
        \bottomrule[1.5pt] 
    \end{tabular}
    }
\end{table}

\begin{table}[t]
    \caption{Comparison of PPL.}
    \label{ppl}
    \resizebox{0.65\linewidth}{!}{
    \begin{tabular}{lcccccc}
    \toprule[1.5pt]
    \multirow{2}{*}{Method} & \multicolumn{3}{c}{Dataset} \\
    \cmidrule(lr){2-4}
    & PersonaChat & C4 & IMDB \\
    \midrule
    AC & 112.70 & 157.78 & 213.95 \\
    ADG & 127.24 & 398.31 & 394.51 \\
    ADG+SECC & 125.16 & 390.18 & 431.72 \\
    Discop & 109.34 & 193.18 & 176.15 \\
    STEAD & 128.55 & 94.07 & 291.76 \\
    Ours & \textbf{53.85} & \textbf{43.21} & \textbf{41.52} \\
    \bottomrule[1.5pt]
    \end{tabular}
    }
\end{table}

\begin{table}[t]
    \centering
    \caption{Embedding capacity when $\alpha=5$. The ``Ours (P/C/I)'' column represents the result of the stegos of our method on PersonaChat/C4/IMDB.}
    \resizebox{1\linewidth}{!}{
        \begin{tabular}{lccccccccc}
        \toprule[1.5pt]
        Method & AC & ADG & ADG+SECC & Discop & STEAD & FSBTS & Ours (P/C/I) \\
        \midrule
        $P_E$ & 0.4001 & 0.3852 &	0.4938 & 0.2856 & 0.4204 & {0.0438} & \textbf{0.0038} \\
        $H_2$ & 0.971 &	0.958 &	1 &	0.863 &	0.977 & {0.0259} & \textbf{0.036} \\
        ER (bit/token) & \textbf{2.47} &	{1.36} &	0.45 &	0.43 &	0.08 & 0.48 & 0.14/0.08/0.07 \\
        EER (bit/token) & 0.07 &	0.06 &	0.00 &	0.06 &	0.00 &	\textbf{0.36} & {0.13}/0.08/0.07 \\
        \bottomrule[1.5pt]
    \end{tabular}
    }
\label{er-w}
\end{table}

\begin{figure*}[t]
\centering
\includegraphics[width=1\textwidth]{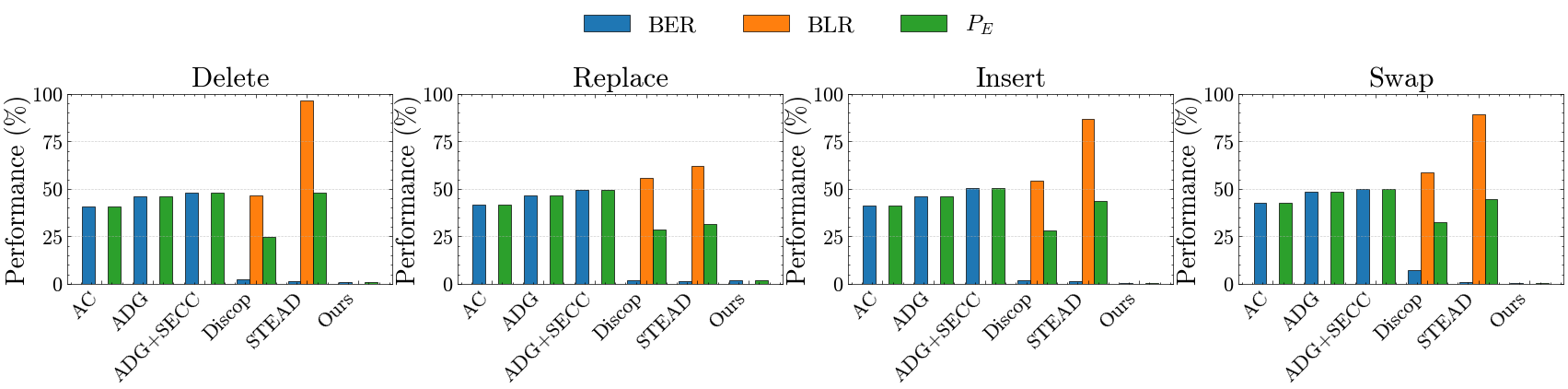}
\caption{The specific robustness performance of different methods under different attack types when $\alpha$=5.}
\label{specific-robust}
\end{figure*}

\noindent{\textbf{Security.}} We use five steganalysis strategies—FCN~\cite{yang2019fast}, R-BiLSTM-C~\cite{niu2019hybrid}, LSTMATT~\cite{zou2020high}, SANet~\cite{xue2023adaptive} and HiDuNet~\cite{peng2023text}—to test security of stegos. The closer the accuracy of steganalysis to 50\%, the better the concealment and the securer the stegos. Since there is no comparable cover text for FSBTS, it is not included in the evaluation.

\subsection{Results}
\begin{figure*}[t]
\centering
\includegraphics[width=1\textwidth,height=4.5cm]{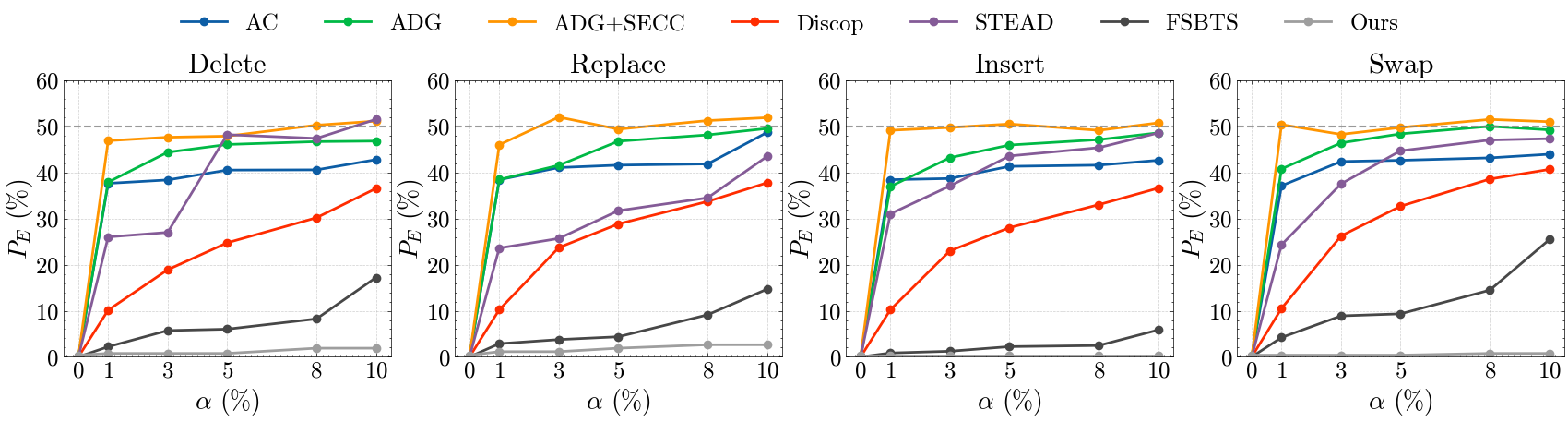}
\caption{Robustness against word-level attacks. $\alpha$ represents the ratio of the attacked words to the total number of words in a sample.}
\label{word-robust}
\end{figure*}

\noindent{\textbf{Robustness.}} Figure~\ref{word-robust} shows the robust performance of different methods under different word-level attack types with different attack intensities. Our method sets the redundancy of LT code to three times. It can be seen that the baseline methods are quite vulnerable to word-level attacks, and their $P_E$ increase as the attacks become stronger, tending to be 50\% of random guess. In contrast, the $P_E$ of our proposed method remains below 2.7\%, achieved A significant improvement compared to the baseline methods.

Figure~\ref{specific-robust} illustrates the performance of various methods in terms of BER, BLR, and $P_E$ under different types of attacks when $\alpha$ = 5. Based on the results, we can thoroughly analyze the primary factors contributing to the extraction errors for each method. First, the $P_E$ of AC and ADG is dominated by BER, as assigning default values to undecodable tokens converts potential loss into extraction errors. Although ADG extracts sufficient blocks for SECC to achieve a 0 BLR, its high inherent symbol error rate causes SECC's error correction to fail, yielding a BER near 50\%. Conversely, our method shares SECC's BLR mechanism but ensures highly accurate symbol extraction via resynchronization and semantic space stability. This successfully averts SECC's decoding failure, maintaining exceptionally low BER and $P_E$ under various attacks. Second, the $P_E$ of Discop and STEAD is predominantly driven by BLR, because they halt decoding at undecodable tokens, converting potential errors into message loss. Notably, Discop's apparent error rate is artificially lowered by its zero-padding strategy, as these padded sequences are excluded from actual BLR calculations.

Table \ref{para-robust} shows the robustness performance of different methods under paraphrase attack. It can be clearly seen that the baseline methods are also vulnerable to sentence-level attack, with Discop and STEAD completely unable to decode the attacked stegos, while our method still has strong robustness and significant advantages.

Among all attacks, the ER is slightly higher for replacement and deletion attacks on our system, as these operations may alter semantic keywords, causing the prediction to fail to be assigned to the correct cluster.

\begin{table}[t]
    \centering
    \caption{Anti-steganalysis comparison. The ADG results are the same whether or not SECC error correction is added.}
    \resizebox{1\linewidth}{!}{
    \begin{tabular}{lccccc}
        \toprule[1.5pt]
        Method  & FCN & R-BiLSTM-C & LSTMATT & SANet & HiDuNet\\
        \hline
        AC     & 65.10\%  & 59.56\% & 59.75\% & 77.58\% & 74.00\%\\
        ADG(+SECC) & 58.30\% & 56.93\% & 53.82\%  & 90.67\% & 96.33\%\\
        Discop  & 52.70\% & 56.90\% & 53.92\% & 90.67\% & 92.71\%\\
        STEAD & 55.56\% & 54.44\% & 52.22\% & \textbf{52.54}\% & 93.33\%\\
        Ours & \textbf{50.54}\% & \textbf{49.60}\% & \textbf{50.63}\% & 53.43\% & \textbf{56.49\%}\\
        \bottomrule[1.5pt]
    \end{tabular}
    }
    \label{security}
\end{table}

\begin{table}[t]
\centering
    \caption{The effect of the number of clusters $k$ on robustness and capacity ($\alpha=5$).}
    \label{k-robust-er}
    \resizebox{0.91\linewidth}{!}{
    \begin{tabular}{lcccccc}
        \toprule[1.5pt]
        $k$  & $P_{ED}$ & $P_{ER}$ & $P_{EI}$ & $P_{ES}$ & $P_{EP}$ & ER\\
        \hline
        2    & 0.0038 & 0.0132 & 0 & 0 & 0 & 0.0299\\
        4    & 0.0132 & 0.0225 & 0 & 0.0019 & 0 & 0.0667 \\
        8    & 0.0188 & 0.0263 & 0.0038 & 0.0038 & 0.0038 & 0.1295 \\
        16   & 0.2876 & 0.1147 & 0.3214 & 0.3195 & 0.3214 & 0.1338 \\
        \bottomrule[1.5pt]
    \end{tabular}}
\end{table}

\noindent{\textbf{Quality of Stegos.}} Table ~\ref{ppl} shows the PPL metrics of different methods on three datasets: PersonaChat~\cite{zhang2018personalizing}, C4~\cite{raffel2020exploring}, and IMDB~\cite{maas2011learning}. By selecting natural sentences directly from the corpus, our method preserves the original linguistic distribution and fluency, achieving significantly lower PPL compared to generative methods.

\noindent{\textbf{Embedding Capacity.}} As shown in Table \ref{er-w}, we test the embedding capacity of different methods under mixed attacks (including all word-level attacks) when $\alpha$=5. Our method sets the redundancy of LT code to three times. Although our method offers no advantage in ER under lossless transmission, its EER in the attack scenario outperforms that of other methods except FSBTS, which employs a larger dataset than ours. In addition to the comparison with the baseline methods, we test the embedding capacity of the proposed method on the same three datasets as mentioned above. It is clear that the ER of the stego texts in PersonaChat is higher, as the average length of its sentences is significantly shorter than that of the other two datasets.

\begin{table}[t]
    \centering
    \caption{Ablation study of HCM. We test the efficiency under different numbers of clustering layers.}
    \label{time}
    \resizebox{1\linewidth}{!}{
    \begin{tabular}{ccccc}
        \toprule[1.5pt]
        \makecell{Number\\of Layers} & \makecell{Encoding\\Rate(s/bit)}  & \makecell{Encoding\\Rate(s/layer)} & \makecell{Decoding\\Rate(s/bit)} & \makecell{Decoding\\Rate(s/layer)}\\
        \hline
        1 & 20.96 & 167.71 & 10.75 & 85.96  \\
        2 & \textbf{1.37} & 5.48 & \textbf{1.17} & 4.67 \\
        3 & 1.52 & 4.61 & 1.49 & 4.47 \\
        4 & 2.18 & \textbf{1.09} & 2.11 & \textbf{1.06} \\
        \bottomrule[1.5pt]
    \end{tabular}
    }
\end{table}

\begin{table}[t]
    \centering
    \caption{Capacity under different redundancy levels for deletion attack ($\alpha=5$).}
    \label{LTR-er}
    \setlength{\tabcolsep}{12pt}
    \resizebox{1\linewidth}{!}{
    \begin{tabular}{lcccc}
        \toprule[1.5pt]
        $R_{LT}$  & $1.5\times$ & $2\times$ & $2.5\times$ & $3\times$\\
        \hline
        ER(bit/token)     & 0.2552 & 0.1925 & 0.1598 & 0.1338 \\
        \bottomrule[1.5pt]
    \end{tabular}
    }
\end{table}

\begin{figure*}[t]
\centering
\includegraphics[width=1\textwidth,height=4.5cm]{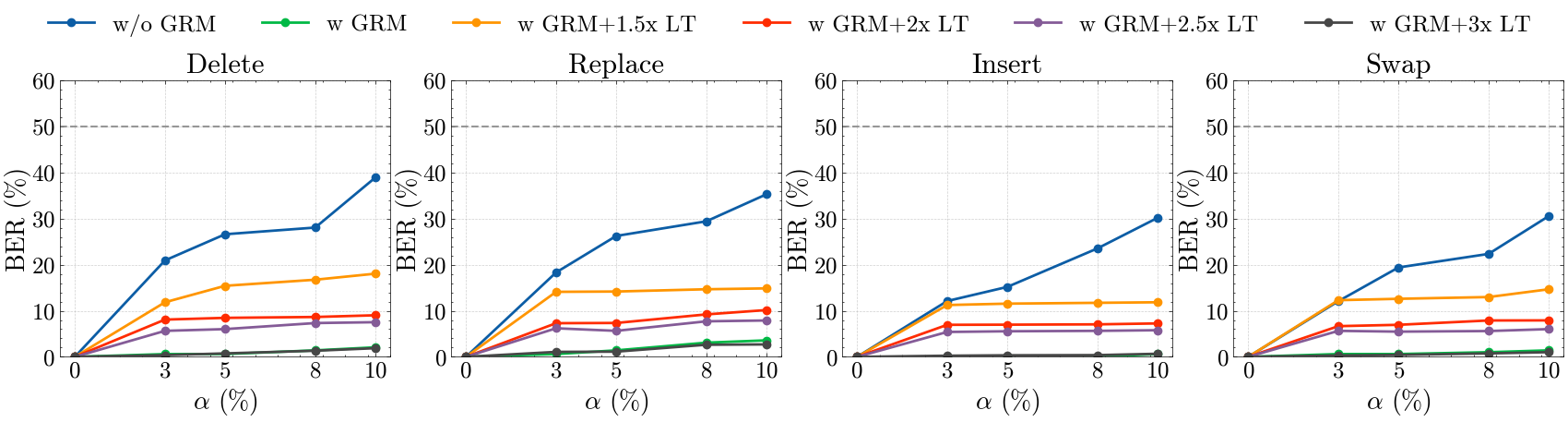}
\caption{Ablation study of GRM. We test the BER of systems without GRM and systems with GRM combined with LT codes at different redundancy levels under various types of attacks of varying intensity.}
\label{GR}
\end{figure*}

\noindent{\textbf{Security.}} Table \ref{security} shows that the detection accuracy of our proposed method is close to 50\%, which shows that the performance of steganalysis methods is not better than random guess when detecting the stegos generated by our method, thus verifying the security of our method.

\subsection{Ablation Study}
\noindent{\textbf{Number of Clusters.}} We test how the system's robustness and embedding capacity vary with different cluster sizes $k$. Columns 2 through 6 represent the $P_E$ values under deletion, replacement, insertion, swap, and paraphrase attacks, respectively. Table \ref{k-robust-er} shows that as $k$ increases, $P_E$ exhibits an upward trend, indicating that this reduces the size of the subspace, thereby making extraction errors more likely to occur. In addition, an increase in $k$ also increases the codable entropy and improves the embedding rate (ER).

\noindent{\textbf{Hierarchical Clustering Mechanism (HCM).}} We test the time consumption with different number of levels. Table ~\ref{time} shows that, compared with the single-layer flat structure, the introduction of HCM greatly reduces the single clustering time and achieves an order of huge improvement in efficiency. When the number is 2, the system reaches the efficiency inflection point; When the number continue to increase, time spent on each level falls, but the cross-layer scheduling overhead caused by multi-layer begins to exceed the benefit, resulting in a rebound in the overall time per bit.

\noindent{\textbf{Global Resynchronization Mechanism (GRM).}} As illustrated in Figure ~\ref{GR}, the performance drops significantly without GRM module. The system with GRM achieves an extremely low BER, but the BER increases slightly when LT codes are added. This is because GRM converts error bits into lost bits, whereas LT’s redundancy coding converts them back into fewer error bits. This verifies the inherent robustness of semantic-space steganography and confirms that bit-slippage, rather than semantic distortion, is the predominant factor behind decoding failure in the baseline system. In addition, the BER decreases consistently as the redundancy of LT codes increases. While 2× redundancy keeps the BER below 10\%, it still exhibits an error floor. In contrast, the 3× configuration achieves near-perfect reconstruction across all attack types and intensities. The system maintains consistent robustness across four types of attacks.

Furthermore, as shown in Table \ref{LTR-er}, as the redundancy of the LT code increases, the amount of actual secret messages embedded decreases, and the ER shows a downward trend. This illustrates the trade-off between robustness and embedding capacity.

\section{Conclusion}

This paper proposes a robust coverless linguistic steganography framework that operates in the sentence embedding space rather than the fragile token space. By leveraging the semantic invariance of sentence representations, our method inherently absorbs word- and sentence-level textual perturbations. We introduce HCM to efficiently partition the semantic space and GRM to resolve the critical bit-slippage problem through LT codes. Experimental results demonstrate that our approach achieves substantial improvements in robustness, while maintaining effective embedding capacity, superior linguistic quality, and strong security against steganalysis.



\begin{thebibliography}{44}


\ifx \showCODEN    \undefined \def \showCODEN     #1{\unskip}     \fi
\ifx \showISBNx    \undefined \def \showISBNx     #1{\unskip}     \fi
\ifx \showISBNxiii \undefined \def \showISBNxiii  #1{\unskip}     \fi
\ifx \showISSN     \undefined \def \showISSN      #1{\unskip}     \fi
\ifx \showLCCN     \undefined \def \showLCCN      #1{\unskip}     \fi
\ifx \shownote     \undefined \def \shownote      #1{#1}          \fi
\ifx \showarticletitle \undefined \def \showarticletitle #1{#1}   \fi
\ifx \showURL      \undefined \def \showURL       {\relax}        \fi
\providecommand\bibfield[2]{#2}
\providecommand\bibinfo[2]{#2}
\providecommand\natexlab[1]{#1}
\providecommand\showeprint[2][]{arXiv:#2}

\bibitem[Chen et~al\mbox{.}(2024)]%
        {chen2024bge}
\bibfield{author}{\bibinfo{person}{Jianlv Chen}, \bibinfo{person}{Shitao Xiao}, \bibinfo{person}{Peitian Zhang}, \bibinfo{person}{Kun Luo}, \bibinfo{person}{Defu Lian}, {and} \bibinfo{person}{Zheng Liu}.} \bibinfo{year}{2024}\natexlab{}.
\newblock \showarticletitle{Bge m3-embedding: Multi-lingual, multi-functionality, multi-granularity text embeddings through self-knowledge distillation}.
\newblock \bibinfo{journal}{\emph{arXiv preprint arXiv:2402.03216}} \bibinfo{volume}{4}, \bibinfo{number}{5} (\bibinfo{year}{2024}).
\newblock


\bibitem[Chen et~al\mbox{.}(2019)]%
        {chen2019adaptive}
\bibfield{author}{\bibinfo{person}{Yanli Chen}, \bibinfo{person}{Hongxia Wang}, \bibinfo{person}{Hanzhou Wu}, \bibinfo{person}{Zhiqiang Wu}, \bibinfo{person}{Tao Li}, {and} \bibinfo{person}{Asad Malik}.} \bibinfo{year}{2019}\natexlab{}.
\newblock \showarticletitle{Adaptive video data hiding through cost assignment and STCs}.
\newblock \bibinfo{journal}{\emph{IEEE Transactions on Dependable and Secure Computing}} \bibinfo{volume}{18}, \bibinfo{number}{3} (\bibinfo{year}{2019}), \bibinfo{pages}{1320--1335}.
\newblock


\bibitem[Damodaran(2021)]%
        {prithivida2021parrot}
\bibfield{author}{\bibinfo{person}{Prithiviraj Damodaran}.} \bibinfo{year}{2021}\natexlab{}.
\newblock \bibinfo{title}{Parrot: Paraphrase generation for NLU.}
\newblock


\bibitem[Ding et~al\mbox{.}(2023)]%
        {ding2023discop}
\bibfield{author}{\bibinfo{person}{Jinyang Ding}, \bibinfo{person}{Kejiang Chen}, \bibinfo{person}{Yaofei Wang}, \bibinfo{person}{Na Zhao}, \bibinfo{person}{Weiming Zhang}, {and} \bibinfo{person}{Nenghai Yu}.} \bibinfo{year}{2023}\natexlab{}.
\newblock \showarticletitle{Discop: Provably secure steganography in practice based on" distribution copies"}. In \bibinfo{booktitle}{\emph{2023 IEEE Symposium on Security and Privacy (SP)}}. IEEE, \bibinfo{pages}{2238--2255}.
\newblock


\bibitem[Guo et~al\mbox{.}(2025)]%
        {guo2025secc}
\bibfield{author}{\bibinfo{person}{Yuzhe Guo}, \bibinfo{person}{Zhongliang Yang}, \bibinfo{person}{Zhuang Wang}, \bibinfo{person}{Zhili Zhou}, {and} \bibinfo{person}{Linna Zhou}.} \bibinfo{year}{2025}\natexlab{}.
\newblock \showarticletitle{SECC-Stega: Generative Linguistic Steganographic Framework Based on Error Correcting Codes}. In \bibinfo{booktitle}{\emph{ICASSP 2025-2025 IEEE International Conference on Acoustics, Speech and Signal Processing (ICASSP)}}. IEEE, \bibinfo{pages}{1--5}.
\newblock


\bibitem[Hu et~al\mbox{.}(2020)]%
        {hu2020mm}
\bibfield{author}{\bibinfo{person}{Yuting Hu}, \bibinfo{person}{Haoyun Li}, \bibinfo{person}{Jianni Song}, {and} \bibinfo{person}{Yongfeng Huang}.} \bibinfo{year}{2020}\natexlab{}.
\newblock \showarticletitle{MM-stega: multi-modal steganography based on text-image matching}. In \bibinfo{booktitle}{\emph{International Conference on Artificial Intelligence and Security}}. Springer, \bibinfo{pages}{313--325}.
\newblock


\bibitem[Kaptchuk et~al\mbox{.}(2021)]%
        {kaptchuk2021meteor}
\bibfield{author}{\bibinfo{person}{Gabriel Kaptchuk}, \bibinfo{person}{Tushar~M Jois}, \bibinfo{person}{Matthew Green}, {and} \bibinfo{person}{Aviel~D Rubin}.} \bibinfo{year}{2021}\natexlab{}.
\newblock \showarticletitle{Meteor: Cryptographically secure steganography for realistic distributions}. In \bibinfo{booktitle}{\emph{Proceedings of the 2021 ACM SIGSAC Conference on Computer and Communications Security}}. \bibinfo{pages}{1529--1548}.
\newblock


\bibitem[Karpukhin et~al\mbox{.}(2020)]%
        {karpukhin2020dense}
\bibfield{author}{\bibinfo{person}{Vladimir Karpukhin}, \bibinfo{person}{Barlas Oguz}, \bibinfo{person}{Sewon Min}, \bibinfo{person}{Patrick~SH Lewis}, \bibinfo{person}{Ledell Wu}, \bibinfo{person}{Sergey Edunov}, \bibinfo{person}{Danqi Chen}, {and} \bibinfo{person}{Wen-tau Yih}.} \bibinfo{year}{2020}\natexlab{}.
\newblock \showarticletitle{Dense Passage Retrieval for Open-Domain Question Answering.}. In \bibinfo{booktitle}{\emph{EMNLP (1)}}. \bibinfo{pages}{6769--6781}.
\newblock


\bibitem[Long and Liu(2018)]%
        {long2018text}
\bibfield{author}{\bibinfo{person}{Yi Long} {and} \bibinfo{person}{Yuling Liu}.} \bibinfo{year}{2018}\natexlab{}.
\newblock \showarticletitle{Text coverless information hiding based on word2vec}. In \bibinfo{booktitle}{\emph{International Conference on Cloud Computing and Security}}. Springer, \bibinfo{pages}{463--472}.
\newblock


\bibitem[Luby(2002)]%
        {luby2002lt}
\bibfield{author}{\bibinfo{person}{Michael Luby}.} \bibinfo{year}{2002}\natexlab{}.
\newblock \showarticletitle{LT codes}. In \bibinfo{booktitle}{\emph{The 43rd Annual IEEE Symposium on Foundations of Computer Science, 2002. Proceedings.}} IEEE Computer Society, \bibinfo{pages}{271--271}.
\newblock


\bibitem[Maas et~al\mbox{.}(2011)]%
        {maas2011learning}
\bibfield{author}{\bibinfo{person}{Andrew Maas}, \bibinfo{person}{Raymond~E Daly}, \bibinfo{person}{Peter~T Pham}, \bibinfo{person}{Dan Huang}, \bibinfo{person}{Andrew~Y Ng}, {and} \bibinfo{person}{Christopher Potts}.} \bibinfo{year}{2011}\natexlab{}.
\newblock \showarticletitle{Learning word vectors for sentiment analysis}. In \bibinfo{booktitle}{\emph{Proceedings of the 49th annual meeting of the association for computational linguistics: Human language technologies}}. \bibinfo{pages}{142--150}.
\newblock


\bibitem[Mikolov et~al\mbox{.}(2013)]%
        {mikolov2013efficient}
\bibfield{author}{\bibinfo{person}{Tomas Mikolov}, \bibinfo{person}{Kai Chen}, \bibinfo{person}{Greg Corrado}, {and} \bibinfo{person}{Jeffrey Dean}.} \bibinfo{year}{2013}\natexlab{}.
\newblock \showarticletitle{Efficient estimation of word representations in vector space}.
\newblock \bibinfo{journal}{\emph{arXiv preprint arXiv:1301.3781}} (\bibinfo{year}{2013}).
\newblock


\bibitem[Ni et~al\mbox{.}(2022)]%
        {ni2022sentence}
\bibfield{author}{\bibinfo{person}{Jianmo Ni}, \bibinfo{person}{Gustavo~Hernandez Abrego}, \bibinfo{person}{Noah Constant}, \bibinfo{person}{Ji Ma}, \bibinfo{person}{Keith Hall}, \bibinfo{person}{Daniel Cer}, {and} \bibinfo{person}{Yinfei Yang}.} \bibinfo{year}{2022}\natexlab{}.
\newblock \showarticletitle{Sentence-t5: Scalable sentence encoders from pre-trained text-to-text models}. In \bibinfo{booktitle}{\emph{Findings of the association for computational linguistics: ACL 2022}}. \bibinfo{pages}{1864--1874}.
\newblock


\bibitem[Niu et~al\mbox{.}(2019)]%
        {niu2019hybrid}
\bibfield{author}{\bibinfo{person}{Yan Niu}, \bibinfo{person}{Juan Wen}, \bibinfo{person}{Ping Zhong}, {and} \bibinfo{person}{Yiming Xue}.} \bibinfo{year}{2019}\natexlab{}.
\newblock \showarticletitle{A hybrid R-BILSTM-C neural network based text steganalysis}.
\newblock \bibinfo{journal}{\emph{IEEE Signal Processing Letters}} \bibinfo{volume}{26}, \bibinfo{number}{12} (\bibinfo{year}{2019}), \bibinfo{pages}{1907--1911}.
\newblock


\bibitem[Pan et~al\mbox{.}(2025)]%
        {pan2025rethinking}
\bibfield{author}{\bibinfo{person}{Chao Pan}, \bibinfo{person}{Donghui Hu}, \bibinfo{person}{Yaofei Wang}, \bibinfo{person}{Kejiang Chen}, \bibinfo{person}{Yinyin Peng}, \bibinfo{person}{Xianjin Rong}, \bibinfo{person}{Chen Gu}, {and} \bibinfo{person}{Meng Li}.} \bibinfo{year}{2025}\natexlab{}.
\newblock \showarticletitle{Rethinking Prefix-Based Steganography for Enhanced Security and Efficiency}.
\newblock \bibinfo{journal}{\emph{IEEE Transactions on Information Forensics and Security}} (\bibinfo{year}{2025}).
\newblock


\bibitem[Pang et~al\mbox{.}(2025)]%
        {pang2025winstega}
\bibfield{author}{\bibinfo{person}{Kaiyi Pang}, \bibinfo{person}{Minhao Bai}, \bibinfo{person}{Jinshuai Yang}, \bibinfo{person}{Wei-Qiang Zhang}, \bibinfo{person}{Minghu Jiang}, {and} \bibinfo{person}{Yongfeng Huang}.} \bibinfo{year}{2025}\natexlab{}.
\newblock \showarticletitle{WinStega: An Adaptive Robust Enhancement Framework for Generative Linguistic Steganography}. In \bibinfo{booktitle}{\emph{ICASSP 2025-2025 IEEE International Conference on Acoustics, Speech and Signal Processing (ICASSP)}}. IEEE, \bibinfo{pages}{1--5}.
\newblock


\bibitem[Peng et~al\mbox{.}(2023)]%
        {peng2023text}
\bibfield{author}{\bibinfo{person}{Wanli Peng}, \bibinfo{person}{Sheng Li}, \bibinfo{person}{Zhenxing Qian}, {and} \bibinfo{person}{Xinpeng Zhang}.} \bibinfo{year}{2023}\natexlab{}.
\newblock \showarticletitle{Text steganalysis based on hierarchical supervised learning and dual attention mechanism}.
\newblock \bibinfo{journal}{\emph{IEEE/ACM Transactions on Audio, Speech, and Language Processing}}  \bibinfo{volume}{31} (\bibinfo{year}{2023}), \bibinfo{pages}{3513--3526}.
\newblock


\bibitem[Qi et~al\mbox{.}(2024)]%
        {qi2024provably}
\bibfield{author}{\bibinfo{person}{Yuang Qi}, \bibinfo{person}{Kejiang Chen}, \bibinfo{person}{Kai Zeng}, \bibinfo{person}{Weiming Zhang}, {and} \bibinfo{person}{Nenghai Yu}.} \bibinfo{year}{2024}\natexlab{}.
\newblock \showarticletitle{Provably secure disambiguating neural linguistic steganography}.
\newblock \bibinfo{journal}{\emph{IEEE Transactions on Dependable and Secure Computing}} (\bibinfo{year}{2024}).
\newblock


\bibitem[Qi et~al\mbox{.}(2025)]%
        {qi2025stead}
\bibfield{author}{\bibinfo{person}{Yuang Qi}, \bibinfo{person}{Na Zhao}, \bibinfo{person}{Qiyi Yao}, \bibinfo{person}{Benlong Wu}, \bibinfo{person}{Weiming Zhang}, \bibinfo{person}{Nenghai Yu}, {and} \bibinfo{person}{Kejiang Chen}.} \bibinfo{year}{2025}\natexlab{}.
\newblock \showarticletitle{STEAD: Robust Provably Secure Linguistic Steganography with Diffusion Language Model}. In \bibinfo{booktitle}{\emph{The Thirty-ninth Annual Conference on Neural Information Processing Systems}}.
\newblock


\bibitem[Radford et~al\mbox{.}(2019)]%
        {radford2019language}
\bibfield{author}{\bibinfo{person}{Alec Radford}, \bibinfo{person}{Jeffrey Wu}, \bibinfo{person}{Rewon Child}, \bibinfo{person}{David Luan}, \bibinfo{person}{Dario Amodei}, \bibinfo{person}{Ilya Sutskever}, {et~al\mbox{.}}} \bibinfo{year}{2019}\natexlab{}.
\newblock \showarticletitle{Language models are unsupervised multitask learners}.
\newblock \bibinfo{journal}{\emph{OpenAI blog}} \bibinfo{volume}{1}, \bibinfo{number}{8} (\bibinfo{year}{2019}), \bibinfo{pages}{9}.
\newblock


\bibitem[Raffel et~al\mbox{.}(2020)]%
        {raffel2020exploring}
\bibfield{author}{\bibinfo{person}{Colin Raffel}, \bibinfo{person}{Noam Shazeer}, \bibinfo{person}{Adam Roberts}, \bibinfo{person}{Katherine Lee}, \bibinfo{person}{Sharan Narang}, \bibinfo{person}{Michael Matena}, \bibinfo{person}{Yanqi Zhou}, \bibinfo{person}{Wei Li}, {and} \bibinfo{person}{Peter~J Liu}.} \bibinfo{year}{2020}\natexlab{}.
\newblock \showarticletitle{Exploring the limits of transfer learning with a unified text-to-text transformer}.
\newblock \bibinfo{journal}{\emph{Journal of machine learning research}} \bibinfo{volume}{21}, \bibinfo{number}{140} (\bibinfo{year}{2020}), \bibinfo{pages}{1--67}.
\newblock


\bibitem[Shannon(1948)]%
        {shannon1948mathematical}
\bibfield{author}{\bibinfo{person}{Claude~E Shannon}.} \bibinfo{year}{1948}\natexlab{}.
\newblock \showarticletitle{A mathematical theory of communication}.
\newblock \bibinfo{journal}{\emph{The Bell system technical journal}} \bibinfo{volume}{27}, \bibinfo{number}{3} (\bibinfo{year}{1948}), \bibinfo{pages}{379--423}.
\newblock


\bibitem[Su et~al\mbox{.}(2021)]%
        {su2021whitening}
\bibfield{author}{\bibinfo{person}{Jianlin Su}, \bibinfo{person}{Jiarun Cao}, \bibinfo{person}{Weijie Liu}, {and} \bibinfo{person}{Yangyiwen Ou}.} \bibinfo{year}{2021}\natexlab{}.
\newblock \showarticletitle{Whitening sentence representations for better semantics and faster retrieval}.
\newblock \bibinfo{journal}{\emph{arXiv preprint arXiv:2103.15316}} (\bibinfo{year}{2021}).
\newblock


\bibitem[Touvron et~al\mbox{.}(2023)]%
        {touvron2023llama}
\bibfield{author}{\bibinfo{person}{Hugo Touvron}, \bibinfo{person}{Thibaut Lavril}, \bibinfo{person}{Gautier Izacard}, \bibinfo{person}{Xavier Martinet}, \bibinfo{person}{Marie-Anne Lachaux}, \bibinfo{person}{Timoth{\'e}e Lacroix}, \bibinfo{person}{Baptiste Rozi{\`e}re}, \bibinfo{person}{Naman Goyal}, \bibinfo{person}{Eric Hambro}, \bibinfo{person}{Faisal Azhar}, {et~al\mbox{.}}} \bibinfo{year}{2023}\natexlab{}.
\newblock \showarticletitle{Llama: Open and efficient foundation language models}.
\newblock \bibinfo{journal}{\emph{arXiv preprint arXiv:2302.13971}} (\bibinfo{year}{2023}).
\newblock


\bibitem[Wang and Gao(2019)]%
        {wang2019coverless}
\bibfield{author}{\bibinfo{person}{Kaixi Wang} {and} \bibinfo{person}{Quansheng Gao}.} \bibinfo{year}{2019}\natexlab{}.
\newblock \showarticletitle{A coverless plain text steganography based on character features}.
\newblock \bibinfo{journal}{\emph{IEEE Access}}  \bibinfo{volume}{7} (\bibinfo{year}{2019}), \bibinfo{pages}{95665--95676}.
\newblock


\bibitem[Wang et~al\mbox{.}(2025)]%
        {wang2025sparsamp}
\bibfield{author}{\bibinfo{person}{Yaofei Wang}, \bibinfo{person}{Gang Pei}, \bibinfo{person}{Kejiang Chen}, \bibinfo{person}{Jinyang Ding}, \bibinfo{person}{Chao Pan}, \bibinfo{person}{Weilong Pang}, \bibinfo{person}{Donghui Hu}, {and} \bibinfo{person}{Weiming Zhang}.} \bibinfo{year}{2025}\natexlab{}.
\newblock \showarticletitle{$\{$SparSamp$\}$: Efficient Provably Secure Steganography Based on Sparse Sampling}. In \bibinfo{booktitle}{\emph{34th USENIX Security Symposium (USENIX Security 25)}}. \bibinfo{pages}{6817--6835}.
\newblock


\bibitem[Wilson and Ker(2016)]%
        {wilson2016avoiding}
\bibfield{author}{\bibinfo{person}{Alex Wilson} {and} \bibinfo{person}{Andrew~D Ker}.} \bibinfo{year}{2016}\natexlab{}.
\newblock \showarticletitle{Avoiding detection on twitter: embedding strategies for linguistic steganography}.
\newblock \bibinfo{journal}{\emph{Electronic Imaging}}  \bibinfo{volume}{28} (\bibinfo{year}{2016}), \bibinfo{pages}{1--9}.
\newblock


\bibitem[Wu et~al\mbox{.}(2016)]%
        {wu2016separable}
\bibfield{author}{\bibinfo{person}{Han-Zhou Wu}, \bibinfo{person}{Yun-Qing Shi}, \bibinfo{person}{Hong-Xia Wang}, {and} \bibinfo{person}{Lin-Na Zhou}.} \bibinfo{year}{2016}\natexlab{}.
\newblock \showarticletitle{Separable reversible data hiding for encrypted palette images with color partitioning and flipping verification}.
\newblock \bibinfo{journal}{\emph{IEEE transactions on circuits and systems for video technology}} \bibinfo{volume}{27}, \bibinfo{number}{8} (\bibinfo{year}{2016}), \bibinfo{pages}{1620--1631}.
\newblock


\bibitem[Wu et~al\mbox{.}(2020)]%
        {wu2020audio}
\bibfield{author}{\bibinfo{person}{Junqi Wu}, \bibinfo{person}{Bolin Chen}, \bibinfo{person}{Weiqi Luo}, {and} \bibinfo{person}{Yanmei Fang}.} \bibinfo{year}{2020}\natexlab{}.
\newblock \showarticletitle{Audio steganography based on iterative adversarial attacks against convolutional neural networks}.
\newblock \bibinfo{journal}{\emph{IEEE transactions on information forensics and security}}  \bibinfo{volume}{15} (\bibinfo{year}{2020}), \bibinfo{pages}{2282--2294}.
\newblock


\bibitem[Wu et~al\mbox{.}(2024)]%
        {wu2024gtsd}
\bibfield{author}{\bibinfo{person}{Zhengxian Wu}, \bibinfo{person}{Juan Wen}, \bibinfo{person}{Yiming Xue}, \bibinfo{person}{Ziwei Zhang}, {and} \bibinfo{person}{Yinghan Zhou}.} \bibinfo{year}{2024}\natexlab{}.
\newblock \showarticletitle{GTSD: Generative Text Steganography Based on Diffusion Model}. In \bibinfo{booktitle}{\emph{International Conference on Neural Information Processing}}. Springer, \bibinfo{pages}{168--183}.
\newblock


\bibitem[Xue et~al\mbox{.}(2023)]%
        {xue2023adaptive}
\bibfield{author}{\bibinfo{person}{Yiming Xue}, \bibinfo{person}{Jiaxuan Wu}, \bibinfo{person}{Ronghua Ji}, \bibinfo{person}{Ping Zhong}, \bibinfo{person}{Juan Wen}, {and} \bibinfo{person}{Wanli Peng}.} \bibinfo{year}{2023}\natexlab{}.
\newblock \showarticletitle{Adaptive domain-invariant feature extraction for cross-domain linguistic steganalysis}.
\newblock \bibinfo{journal}{\emph{IEEE Transactions on Information Forensics and Security}}  \bibinfo{volume}{19} (\bibinfo{year}{2023}), \bibinfo{pages}{920--933}.
\newblock


\bibitem[Yang et~al\mbox{.}(2025)]%
        {yang2025novel}
\bibfield{author}{\bibinfo{person}{Jinshuai Yang}, \bibinfo{person}{Minghao Bai}, \bibinfo{person}{Kaiyi Pang}, \bibinfo{person}{Yue Gao}, \bibinfo{person}{Jiajun Zou}, {and} \bibinfo{person}{Yongfeng Huang}.} \bibinfo{year}{2025}\natexlab{}.
\newblock \showarticletitle{A Novel Framework of Semantic-Based Text Steganography}.
\newblock \bibinfo{journal}{\emph{IEEE Transactions on Dependable and Secure Computing}} (\bibinfo{year}{2025}).
\newblock


\bibitem[Yang et~al\mbox{.}(2019)]%
        {yang2019fast}
\bibfield{author}{\bibinfo{person}{Zhongliang Yang}, \bibinfo{person}{Yongfeng Huang}, {and} \bibinfo{person}{Yu-Jin Zhang}.} \bibinfo{year}{2019}\natexlab{}.
\newblock \showarticletitle{A fast and efficient text steganalysis method}.
\newblock \bibinfo{journal}{\emph{IEEE Signal Processing Letters}} \bibinfo{volume}{26}, \bibinfo{number}{4} (\bibinfo{year}{2019}), \bibinfo{pages}{627--631}.
\newblock


\bibitem[Yang et~al\mbox{.}(2018)]%
        {yang2018rnn}
\bibfield{author}{\bibinfo{person}{Zhong-Liang Yang}, \bibinfo{person}{Xiao-Qing Guo}, \bibinfo{person}{Zi-Ming Chen}, \bibinfo{person}{Yong-Feng Huang}, {and} \bibinfo{person}{Yu-Jin Zhang}.} \bibinfo{year}{2018}\natexlab{}.
\newblock \showarticletitle{RNN-stega: Linguistic steganography based on recurrent neural networks}.
\newblock \bibinfo{journal}{\emph{IEEE Transactions on Information Forensics and Security}} \bibinfo{volume}{14}, \bibinfo{number}{5} (\bibinfo{year}{2018}), \bibinfo{pages}{1280--1295}.
\newblock


\bibitem[Ye et~al\mbox{.}(2025)]%
        {ye2025dream}
\bibfield{author}{\bibinfo{person}{Jiacheng Ye}, \bibinfo{person}{Zhihui Xie}, \bibinfo{person}{Lin Zheng}, \bibinfo{person}{Jiahui Gao}, \bibinfo{person}{Zirui Wu}, \bibinfo{person}{Xin Jiang}, \bibinfo{person}{Zhenguo Li}, {and} \bibinfo{person}{Lingpeng Kong}.} \bibinfo{year}{2025}\natexlab{}.
\newblock \showarticletitle{Dream 7b: Diffusion large language models}.
\newblock \bibinfo{journal}{\emph{arXiv preprint arXiv:2508.15487}} (\bibinfo{year}{2025}).
\newblock


\bibitem[Yi et~al\mbox{.}(2022)]%
        {yi2022alisa}
\bibfield{author}{\bibinfo{person}{Biao Yi}, \bibinfo{person}{Hanzhou Wu}, \bibinfo{person}{Guorui Feng}, {and} \bibinfo{person}{Xinpeng Zhang}.} \bibinfo{year}{2022}\natexlab{}.
\newblock \showarticletitle{ALiSa: Acrostic linguistic steganography based on BERT and Gibbs sampling}.
\newblock \bibinfo{journal}{\emph{IEEE Signal Processing Letters}}  \bibinfo{volume}{29} (\bibinfo{year}{2022}), \bibinfo{pages}{687--691}.
\newblock


\bibitem[Zhang et~al\mbox{.}(2017)]%
        {zhang2017coverless}
\bibfield{author}{\bibinfo{person}{Jianjun Zhang}, \bibinfo{person}{Yicheng Xie}, \bibinfo{person}{Lucai Wang}, {and} \bibinfo{person}{Haijun Lin}.} \bibinfo{year}{2017}\natexlab{}.
\newblock \showarticletitle{Coverless text information hiding method using the frequent words distance}. In \bibinfo{booktitle}{\emph{International Conference on Cloud Computing and Security}}. Springer, \bibinfo{pages}{121--132}.
\newblock


\bibitem[Zhang et~al\mbox{.}(2021a)]%
        {zhang2021alicg}
\bibfield{author}{\bibinfo{person}{Ningyu Zhang}, \bibinfo{person}{Qianghuai Jia}, \bibinfo{person}{Shumin Deng}, \bibinfo{person}{Xiang Chen}, \bibinfo{person}{Hongbin Ye}, \bibinfo{person}{Hui Chen}, \bibinfo{person}{Huaixiao Tou}, \bibinfo{person}{Gang Huang}, \bibinfo{person}{Zhao Wang}, \bibinfo{person}{Nengwei Hua}, {et~al\mbox{.}}} \bibinfo{year}{2021}\natexlab{a}.
\newblock \showarticletitle{Alicg: Fine-grained and evolvable conceptual graph construction for semantic search at alibaba}. In \bibinfo{booktitle}{\emph{Proceedings of the 27th ACM SIGKDD conference on knowledge discovery \& data mining}}. \bibinfo{pages}{3895--3905}.
\newblock


\bibitem[Zhang et~al\mbox{.}(2018)]%
        {zhang2018personalizing}
\bibfield{author}{\bibinfo{person}{Saizheng Zhang}, \bibinfo{person}{Emily Dinan}, \bibinfo{person}{Jack Urbanek}, \bibinfo{person}{Arthur Szlam}, \bibinfo{person}{Douwe Kiela}, {and} \bibinfo{person}{Jason Weston}.} \bibinfo{year}{2018}\natexlab{}.
\newblock \showarticletitle{Personalizing dialogue agents: I have a dog, do you have pets too?}
\newblock \bibinfo{journal}{\emph{arXiv preprint arXiv:1801.07243}} (\bibinfo{year}{2018}).
\newblock


\bibitem[Zhang et~al\mbox{.}(2021b)]%
        {zhang2021provably}
\bibfield{author}{\bibinfo{person}{Siyu Zhang}, \bibinfo{person}{Zhongliang Yang}, \bibinfo{person}{Jinshuai Yang}, {and} \bibinfo{person}{Yongfeng Huang}.} \bibinfo{year}{2021}\natexlab{b}.
\newblock \showarticletitle{Provably secure generative linguistic steganography}.
\newblock \bibinfo{journal}{\emph{arXiv preprint arXiv:2106.02011}} (\bibinfo{year}{2021}).
\newblock


\bibitem[Zhang et~al\mbox{.}(2020)]%
        {zhang2020dialogpt}
\bibfield{author}{\bibinfo{person}{Yizhe Zhang}, \bibinfo{person}{Siqi Sun}, \bibinfo{person}{Michel Galley}, \bibinfo{person}{Yen-Chun Chen}, \bibinfo{person}{Chris Brockett}, \bibinfo{person}{Xiang Gao}, \bibinfo{person}{Jianfeng Gao}, \bibinfo{person}{Jingjing Liu}, {and} \bibinfo{person}{William~B Dolan}.} \bibinfo{year}{2020}\natexlab{}.
\newblock \showarticletitle{Dialogpt: Large-scale generative pre-training for conversational response generation}. In \bibinfo{booktitle}{\emph{Proceedings of the 58th annual meeting of the association for computational linguistics: system demonstrations}}. \bibinfo{pages}{270--278}.
\newblock


\bibitem[Zhou et~al\mbox{.}(2021)]%
        {zhou2021linguistic}
\bibfield{author}{\bibinfo{person}{Xuejing Zhou}, \bibinfo{person}{Wanli Peng}, \bibinfo{person}{Boya Yang}, \bibinfo{person}{Juan Wen}, \bibinfo{person}{Yiming Xue}, {and} \bibinfo{person}{Ping Zhong}.} \bibinfo{year}{2021}\natexlab{}.
\newblock \showarticletitle{Linguistic steganography based on adaptive probability distribution}.
\newblock \bibinfo{journal}{\emph{IEEE Transactions on Dependable and Secure Computing}} \bibinfo{volume}{19}, \bibinfo{number}{5} (\bibinfo{year}{2021}), \bibinfo{pages}{2982--2997}.
\newblock


\bibitem[Ziegler et~al\mbox{.}(2019)]%
        {ziegler2019neural}
\bibfield{author}{\bibinfo{person}{Zachary Ziegler}, \bibinfo{person}{Yuntian Deng}, {and} \bibinfo{person}{Alexander~M Rush}.} \bibinfo{year}{2019}\natexlab{}.
\newblock \showarticletitle{Neural linguistic steganography}. In \bibinfo{booktitle}{\emph{Proceedings of the 2019 Conference on Empirical Methods in Natural Language Processing and the 9th International Joint Conference on Natural Language Processing (EMNLP-IJCNLP)}}. \bibinfo{pages}{1210--1215}.
\newblock


\bibitem[Zou et~al\mbox{.}(2020)]%
        {zou2020high}
\bibfield{author}{\bibinfo{person}{Jiajun Zou}, \bibinfo{person}{Zhongliang Yang}, \bibinfo{person}{Siyu Zhang}, \bibinfo{person}{Sadaqat~ur Rehman}, {and} \bibinfo{person}{Yongfeng Huang}.} \bibinfo{year}{2020}\natexlab{}.
\newblock \showarticletitle{High-performance linguistic steganalysis, capacity estimation and steganographic positioning}. In \bibinfo{booktitle}{\emph{International Workshop on Digital Watermarking}}. Springer, \bibinfo{pages}{80--93}.
\newblock


\end{thebibliography}

\appendix

\end{document}